\documentclass[runningheads]{llncs}
\usepackage[T1]{fontenc}
\usepackage{graphicx}
\usepackage{xcolor}
\def\beq{\begin{equation}}
\def\eeq{\end{equation}}
\begin{document}
\title{The cross section of inverse  beta decay}
%
%
\author{G.~Ricciardi\inst{1}\orcidID{0000-0002-2352-9033} \and
N.~Vignaroli\inst{2}\orcidID{0000-0002-6711-522X} \and
F.~Vissani\inst{3}\orcidID{0000-0002-8833-5096}}
\authorrunning{G.~Ricciardi et al.}
%
\institute{Dipartimento di Fisica  E. Pancini,   
Universit\`a  di Napoli  Federico II  \\ and INFN,  Naples \\ MSA, Via Cintia, 80126 Naples, Italy \\ 
\email{giulia.ricciardi2@unina.it}\\ \and
Dipartimento di Matematica e Fisica E. De Giorgi, Universit\`a del Salento \\ and INFN,  Lecce\\
Via per Arnesano, 73100 Lecce, Italy \\
\email{natascia.vignaroli@unisalento.it}
\and
INFN Laboratori Nazionali del Gran Sasso,
 L'Aquila, Italy
 \\
\email{francesco.vissani@lngs.infn.it}}

\maketitle              
\begin{abstract}
We discuss the accuracy of the evaluation of the cross section for inverse beta decay at low energies  and its  relevance in the current experimental framework.

\end{abstract}
\section{Introduction}
The interaction process
\begin{equation}
\bar \nu + p \to e^+ + n \qquad \qquad 
\label{inverse}
\end{equation}
is known as  inverse $\beta$ decay\index{beta decay!inverse} (IBD). 
Its cross section was calculated for the first time
by  Bethe and Peierls in 1934~\cite{Bethe:1934qn}, using the  Fermi theory, 
and gave an extremely  small value  ($\sigma \approx 10^{-44}$ cm$^2$ at $E(\bar \nu)=2$ MeV), leading them to believe that the neutrino is an “undetectable particle”. It is kind of ironic that in 1956 the inverse beta decay  was  exactly   the process that allowed the first direct experimental evidence of 
neutrinos~\cite{rc}\footnote{For a brief historical survey see e.g. \cite{Ricciardi:2023tvg}.}. Since then, the IBD has always played a central role in neutrino physics,  on the strength of its frequent occurrence  at  energies below  $\sim 20$ MeV and its wide use in 
water- or hydrocarbon-based detectors, which are relatively cheap materials 
and  rich in target protons.
As a consequence, an accurate estimate of its cross section is essential for the accuracy of the results in several and different  neutrino  experiments.
In this work, we discuss its accurate evaluation \cite{Ricciardi:2022pru} and  relevance in the current experimental framework.

\section{The cross section evaluation}

We summarize the  main steps needed to build the IBD cross section\footnote{For details see Ref. \cite{Ricciardi:2022pru}.}:
\begin{enumerate}
\item write down the  matrix element of the charged weak current between proton and neutron states, which in the  most general form depends on six dimensionless   Lorentz invariant form factors:
\beq
\mathcal{J}_\mu = \,
\bar{u}_n \bigg(  f_1 \gamma_\mu +  g_1 \gamma_\mu \gamma_5 + i f_2 \sigma_{\mu\nu}\frac{q^\nu}{2M} + g_2  \frac{q_\mu}{M} \gamma_5  + f_3  \frac{q_\mu}{M} +  i g_3 \sigma_{\mu\\benu}\frac{q^\nu}{2M} \gamma_5   \bigg) u_p;
\eeq
\item
calculate the tree level amplitude for the  IBD, by using the lepton current and the $W$ propagator. By squaring and averaging, one obtains
the differential cross section
\beq
\frac{d\sigma}{dt} =  \frac{G_F^2 \cos^2\theta_C }{64 \pi (s-m_p^2)^2}   \,  \overline{ |\mathcal{M}^2|}
\eeq
where $G_F$ is the Fermi coupling, the parameters $s = (p_\nu + p_p)^2$, $t = (p_\nu - p_e)^2$ 
 are  Mandelstam variables and $\theta_C$ is the Cabibbo angle
(that is linked to the $u-d$ element of the CKM matrix by the equality $ \cos\theta_C=V_{\mbox{\tiny ud}}$);
\item
 include the radiative corrections 
 at leading order  \cite{Kurylov:2002vj}
\beq
d\sigma(E_\nu,E_e) \to d\sigma(E_\nu,E_e) \left[1 + \frac{\alpha}{\pi} \left(6.00 + \frac{3}{2}\log \frac{m_p}{2 E_e} + 1.2\left( \frac{m_e}{E_e}\right)^{1.5}\right) \right]
\eeq
where $\alpha$ is the fine-structure constant. Next order corrections and other effects such as 
isospin breaking are estimated to be small and  neglected  in the low range of neutrino energies (below  $\sim 10$ MeV).
\end{enumerate}
A new series of modern
calculations was
started by Vogel and Beacom in 1999 \cite{Vogel:1999zy} 
and continued by Strumia and Vissani \cite{Strumia:2003zx} in 2002.
An updated and
accurate assessment
of the uncertainties was
addressed in 2023 by Ricciardi, Vignaroli and Vissani  in 2023 \cite{Ricciardi:2022pru}.

The estimation of theoretical uncertainties  is crucial in experimental 
 measurements aiming at high precision.
In the cross section built as outlined before, the  uncertainties are  reflection of the uncertainties in the knowledge of the form factors and the Cabibbo angle.

The hadronic current $\mathcal{J}_\mu$ includes the vector and axial vector terms with form factors $f_3$ and $g_3$, that transform as  second class currents under G-parity, according  to a classification due to Weinberg \cite{Weinberg:1958ut}.
G-parity is a symmetry for strong interactions, broken by mass differences. Second class currents can be safely neglected at the current level of precision \cite{Ricciardi:2022pru}.

The input parameters which determine the leading 
uncertainties in the cross section evaluation vary with the energy range; they are
\begin{itemize} \item
 the Cabibbo angle  
and the axial coupling $g_1(0)$ (lowest energies);
\item
the axial radius $r_A$ (higher energies).
 \end{itemize}
The magnitude of the  
mixing element $ V_{ud}=\cos \theta_C $  can be derived directly from  the super-allowed   $0^+ \to 0^+$ nuclear
beta decays, which are pure vector transitions. The result is in tension with the one derived indirectly from the CKM unitarity, using the currently accepted values for $|V_{us}|$ and $|V_{ub}|$. That 
has been interpreted as  a failure of unitarity
of around 2$\sigma$ and possibly more \cite{Hardy:2020qwl}.
We believe it is reasonable to assume that this difference merely signals limits to the available interpretations and measurements, and
 we combine the data,  enlarging all errors contributing to the result by the scale factor  
$S=\sqrt{\chi^2/(N-1)}=2.0$
 for a conservative estimation of the uncertainty, in agreement with the PDG suggestion \cite{ParticleDataGroup:2022pth}. We obtain $
 V_{\mbox{\tiny ud}} = 0.9743 (3)$.

There are eight different measurements of the normalized  axial coupling
  $\lambda=-g_1(0)/f_1(0)$ using decay polarized neutrons, with the latest measurement, published in 2019 by the Perkeo III collaboration \cite{Markisch:2018ndu},
being much more precise than the others. 
In a conservative perspective, we include all results,
but enlarging their error by a factor 2 to take into account discrepancies among older and newer results.
That yields the  value $\lambda= 1.2760(5) $, which is within $1\sigma$ from Ref.~\cite{Markisch:2018ndu} and agrees with the global average.

The theoretical relation \cite{Czarnecki:2019mwq}
\begin{equation}
\frac{1}{\tau_{\mbox{\tiny n}}} = \frac{V_{\mbox{\tiny ud}} ^ 2 \ (1+ 3 \lambda ^ 2)} {4906.4 \pm 1.7 \mbox{s}} 
\label{eq:neutron}
\end{equation}
links both $\lambda$ and $ V_{ud}$ to the average lifetime of the neutron $\tau_n$.
By propagating the errors, we find the prediction $ \tau_{\mbox{\tiny n}}{\mbox{\small(SM)}} = 878.38 \pm 0.89 $~s.
Since the neutron decay lifetime $\tau_n$ is measured, Eq. (\ref{eq:neutron})  could help us to improve the inferences on the IBD cross section.
There are two methods of measurement  for the neutron decay lifetime. In  experiments "in-bottle" or "storage", ultra-cold neutrons are trapped and, after a holding time $T$, they are released and  the number $N(T)$ of surviving ultra-cold neutrons is counted. This is repeated for two or more different
holding times, with $T$ ranging from a few minutes up to some fraction of an hour. If no ultra-cold neutrons
are lost, $\tau_n$ is obtained from an exponential fit to $N(T) \propto \exp(-T/\tau_n)$, and no absolute
measurement is required. In "beam" experiments the neutron decay lifetime is directly calculated  by counting decay products (protons
and/or electrons) from a thermal neutron beam, while
simultaneously measuring the neutron beam density\footnote{
 In the past
 years a third method for measuring the neutron life-time has been devised, which uses spacecraft-based neutron detectors to count
the relative neutron flux as a function of altitude \cite{Wilson:2020lfr}. Its  precision is not yet competitive.}.
There is a discrepancy among these results.   In
particular the value reported by the most precise beam
experiment conducted at the NIST Center for Neutron
Research \cite{Yue:2013qrc,Wietfeldt:2022tma} is 8.9 s (3.9$\sigma$) higher than the average  storage value.
This discrepancy  between the two set of $\tau_n$ 
 measurements has been widely discussed in literature, without reaching general consensus.
Since the values of  $V_{ud}$ and $\lambda$ 
 are  consistent with the  values from the storage approach, and incompatible with the ones from the beam approach, we   use only the former data set, assuming that the others are affected by a systematic deviation, which is not yet fully understood.
 The observed value of the neutron lifetime, together with
the SM prediction, yields the relationship
$
  V_{\mbox {\tiny ud}} = 2.36323 (75) /\sqrt {1+ 3 \lambda ^ 2} 
$.

 In summary, by  propagating the uncertainty factors we find that the cross section is known with
$\delta\sigma_{\bar\nu_e p}/\sigma_{\bar\nu_e p}= 0.1\% $
for low values 
 of electron anti-neutrino energies, which is four times
better than~\cite{Strumia:2003zx}.
 
 At higher energies, the cross section becomes sensible to the momenta dependence of the form factors. This is generally expressed by  phenomenological descriptions as the commonly used dipolar approximation. Since we are mostly concerned with low energy processes, whatever the expression of the full dependence is, we  only need the first terms of its Taylor expansion. This is a rather general description and  lessen the dependence on the phenomenological approximations, which in the most common cases are not even optimised for the energies we are discussing.
Thus we set
 $g_1/g_1(0)=1+ q^2 r_A^2/6$. The current  value of the axial mass  \cite{Bodek:2007ym}
 $M_A=1014\pm 14$ MeV in the dipolar approximation implies
 $r_A^2=0.455\pm 0.013$ fm$^2$; a determination that does not assume the double dipole
 has an error larger of about one order of magnitude
  $r_A^2=0.46\pm 0.12$ fm$^2$. 
  
  By including the uncertainty on $r_A$,
 we find 
$\delta\sigma_{\bar\nu_e p}/\sigma_{\bar\nu_e p}=1.1\% (E_\nu/50\mbox{ MeV})^2$ in the 
region above $\sim 10$ MeV,  a value
   which is 
  3 times larger than in \cite{Strumia:2003zx}. 
  
  While the cross section values are consistent with previous ones, the increase in its uncertainty range
motivates attempts to improve the description of form factors at the energies we are considering.
  
  \section{Conclusions}

The IBD plays an important role in present and future experiments studying neutrinos at  low and intermediate energy ranges.

The detection of geo-neutrinos, whose energies extend up to about 2.5 MeV, is  realised  through IBD.
In essentially all  oscillation studies at reactors, electron antineutrinos are  detected  by interactions with free protons via IBD. 
 Reactor neutrino energy spectra end
at $\sim$10 MeV. Some of supernova experiments  based on water Cherenkov detector.
 are primarily sensitive to IBD interactions, revealing the Cherenkov light of the final-state positron. 
The inverse beta decay of supernova neutrinos can also be observed in scintillation  detectors.
Neutrino fluxes from supernovae go up to 50 MeV. 

  A reliable knowledge of the  cross section of the IBD  is based on a  set of  theoretical concepts and on  measurements of the key parameters.
  We assessed the uncertainty on the cross section, which is relevant to experimental advances and increasingly large statistical samples.
  In order to improve the actual estimates, 
  we  need to address the reason of discrepancy in   $\tau_n$ measurements and the unitarity issue in $|V_{ud}|$. This is particularly significant 
  at the low energies which are relevant e.g. for reactor neutrinos. 
 A priority, expecially for supernova studies, is also to refine the description of the axial form factor in the 100 MeV range, thus decreasing the uncertainty due to $r_A$.

\subsubsection{Acknowledgements} This work was partially supported by INFN research initiative ENP and by the research grant number 2022E2J4RK
``PANTHEON: Perspectives in Astroparticle and Neutrino THEory with Old and New messengers"
under the program PRIN 2022 funded by the Italian Ministero dell’Universit\`a e della Ricerca (MUR) and by
the European Union – Next Generation EU.

%
%
%
%

\end{document}